\documentstyle[twocolumn,psfig,prl,aps]{revtex}

\newcommand{\mvec}[1]{\mbox{\boldmath$#1$\unboldmath}}

\begin{document}
\draft

 \twocolumn[\hsize\textwidth\columnwidth\hsize\csname %
 @twocolumnfalse\endcsname

\title{Chirality tunneling in mesoscopic antiferromagnetic domain walls}

\author{B. A. Ivanov,$^{1}$ A. K. Kolezhuk,$^{1,2}$ and V. E. Kireev$^{1}$}
\address{$^{1}$Institute of Magnetism, National Academy of 
Sciences and Ministry of Education of Ukraine,
252142 Kiev, Ukraine\\
$^{2}$Institut f\"{u}r Theoretische Physik,
Universit\"{a}t Hannover, D-30167 Hannover, Germany}

\date{\today}

\maketitle

\begin{abstract}
We consider a domain wall in the mesoscopic quasi-one-dimensional sample
(wire or stripe) of weakly anisotropic two-sublattice antiferromagnet,
and estimate the probability of tunneling between two domain wall
states with different chirality.  Topological effects forbid tunneling
for the systems with half-integer spin $S$ of magnetic atoms which
consist of odd number of chains $n_{\perp}$.  External magnetic field
yields an additional contribution to the Berry phase, resulting in the
appearance of {\em two\/} different tunnel splittings in any
experimental setup involving a mixture of odd and even $n_{\perp}$,
and in oscillating field dependence of the tunneling rate with the
period proportional to $1/n_{\perp}$.
\end{abstract}

\pacs{75.45.+j, 75.60.Ch, 75.50.Ee}

]

In recent years, there have been much interest to the problem of
quantum spin coherence in mesoscopic magnetic systems, mainly in
nanoparticles \cite{QTM}.  Experimentally, one of the crucial problems
is to prepare an ensemble of particles of highly uniform size: even small fluctuations of size lead to large fluctuations of
the tunneling probability since they contribute to the power of
exponent.  One nice solution is to use high-spin molecular clusters
\cite{Gatteschi+94}; another possible way, proposed in
\cite{ivkol94+GalkinaIvanov95,BraunLoss96}, is to use {\em
topologically nontrivial\/} magnetic structures: domain walls in
quasi-one-dimensional (1d) systems (wires, stripes), vortices
in 2d systems, etc. Such objects have mesoscopic
scale, e.g., in weakly anisotropic magnets with magnetic ions in
$s$-states the domain wall thickness is usually about 100 lattice
constants, and since their shape is determined by the material
constants they are to a high extent identical.

Classically, magnetic domain wall (DW) has certain ``chirality,'' an
internal degree of freedom characterizing the way of rotation of
magnetization inside a DW. Two states with opposite chirality are
completely equivalent in energy (we will not consider magnets without
inversion center where this is not true).  In quantum case there is
generally a nonzero transition amplitude mixing the two states and
lifting the degeneracy\cite{ivkol94+GalkinaIvanov95,BraunLoss96};
that under favorable circumstances this tunnel splitting can
be detected with resonant technique of some kind. It is known that in
antiferromagnets (AF) tunneling is generally more favorable than in
ferromagnets, both in case of fine particles
\cite{BarbaraChud+KriveZasl90} and domain walls \cite{BraunLoss96}.

In this paper we show that in the simplest model of mesoscopic AF with
half-integer spin $S$ of magnetic ions topological effects forbid
chirality tunneling for a DW with odd number $n_{\perp}$ of spins in
its cross-section. We further show that in presence of even weak
external magnetic field this strict ``selection rule'' is relaxed,
which leads to appearance of two different values of tunnel splitting
in any half-integer $S$ sample with weakly fluctuating $n_{\perp}$.
For any $S$, the tunneling amplitude is shown to be an oscillating
function of the field.

\begin{figure}
\mbox{\hspace{6mm}\psfig{figure=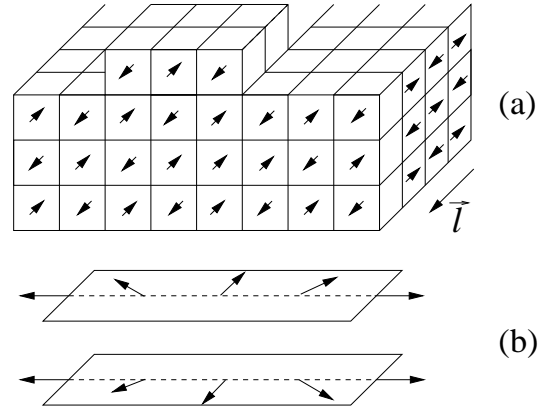,width=70mm,angle=-90.}}
\vspace{3mm}
  \caption{ \label{fig:dw} (a) a schematic picture of the
  cross-section of antiferromagnetic mesoscopic stripe; (b) two domain
  walls with opposite chiralities.}
\end{figure}

{\em Model.\/} Consider a thin quasi-one-dimensional stripe of
two-sublattice weakly anisotropic antiferromagnet, which we for the
sake of simplicity consider as a system of $n_{\perp}$ AF chains of
spin-$S$ magnetic atoms, coupled antiferromagnetically with the same
exchange constant $J$ for any neighboring spins. We assume that
magnetic atoms form a perfect crystal structure on a bipartite
lattice, as shown in Fig.\
\ref{fig:dw}; note that  $n_{\perp}$ can be odd or even
without introducing any frustration. We assume a rhombic anisotropy of
the form
\begin{equation}
  \label{wa}
  w_{a}=\sum_{i}\big[
  K_{1}(S_{i}^{Z})^{2} +K_{2}(S_{i}^{Y})^{2}\big]\,,
\end{equation}
where $i$ labels lattice sites and $K_{1,2}\ll J$ are the anisotropy
constants, $K_{1}>K_{2}>0$, so that $OZ$ is the hard axis and $OX$ is the
easy axis in the easy plane $XY$.

Due to the quasi-1d structure, one can assume that the sublattice
magnetization depends only on the space coordinate $x$ along the wire
(note that $X$ and $x$ axes do not need to coincide). Using the standard
technique \cite{fradkin}, one can obtain the effective Euclidean action of
AF in continuum approximation, which has the form of a well-known
$O(3)$ nonlinear sigma-model:
\begin{eqnarray}
\label{Aeucl}
&&{\cal A}_{E}={1\over4}\, n_{\perp}\hbar S W +i2\pi n_{\perp} S\hbar
(Q +Q_{H}'),\\
&& W[\mvec{l}]=\int d^{2}x \Big\{
(\partial_{\alpha}\mvec{l})(\partial_{\alpha}\mvec{l})
+{1\over\Delta^{2}} \big[(1+\rho)l_{Z}^{2}+l_{Y}^{2}\big]
+\widetilde{w}_{a}\Big\}\,,\nonumber\\
&& Q={1\over4\pi} \int d^{2}x \;
\mvec{l}\cdot(\partial_{1}\mvec{l}\times\partial_{2}\mvec{l})
\,,\nonumber \\
&& Q_{H}'=
{\gamma\over4\pi c}\int d^{2}x\; \mvec{H}\cdot (\mvec{l}\times
\partial_{2}\mvec{l}) \,.\nonumber
\end{eqnarray}
Here $\mvec{l}$ is the unit N\'eel vector, $(x_{1},x_{2})=(x,c\tau)$
is the Euclidean plane, $c=JSaZ_{c}/\hbar$ is the limiting velocity of
spin waves, $Z_{c}$ is the lattice  coordination number, $a$ is
the lattice constant, $\Delta=a(JZ_{c}/4K_{2})^{1/2}\gg a$ is the
characteristic DW thickness, $\rho=K_{1}/K_{2}-1$ is the
rhombicity parameter, $\gamma=g\mu_{B}/\hbar$
is the gyromagnetic ratio, $g$ denotes the Land\'e factor, and
$\mu_{B}$ is the Bohr magneton. The quantity $\widetilde{w}_a(\mvec{l})=(\gamma/
c)^{2}(\mvec{H}\cdot \mvec{l})^{2}$ describes effective renormalization
of the anisotropy induced by the field.  In (\ref{Aeucl}), the term
proportional to $Q$ is the so-called topological term originating from
the sum of Berry phases \cite{Berry84} of individual spins, $Q$ being
the homotopical index of mapping of the $(x_{1},x_{2})$ plane onto the
sphere $\mvec{l}^{2}=1$, and $Q_{H}'$ is the contribution from magnetic
field.

A static DW solution $\mvec{l}_{0}(x)$ corresponds to the
rotation of vector $\mvec{l}$ in the easy plane $XY$:
\begin{equation}
  \label{kink}
  l_{0X}=\sigma'\tanh(x/\Delta),\;\;
   l_{0Y}=\sigma/\cosh(x/\Delta),\;\;
  l_{0Z}=0\,, 
\end{equation}
where $\sigma,\sigma'=\pm1$. The quantity $\sigma'$ is the
``topological charge'' of the DW, and the chirality $\sigma$
determines the sign of $\mvec{l}$ projection onto the ``intermediate''
axis $OY$. Two states with $\sigma=\pm1$ are  equivalent in energy;
change of $\sigma$ describes reorientation of the macroscopic
number of spins $N_{DW}\sim\Delta/a\gg1$, typically
$N_{DW}\sim 70\div100$.

\begin{figure}
\mbox{\hspace{6mm}\psfig{figure=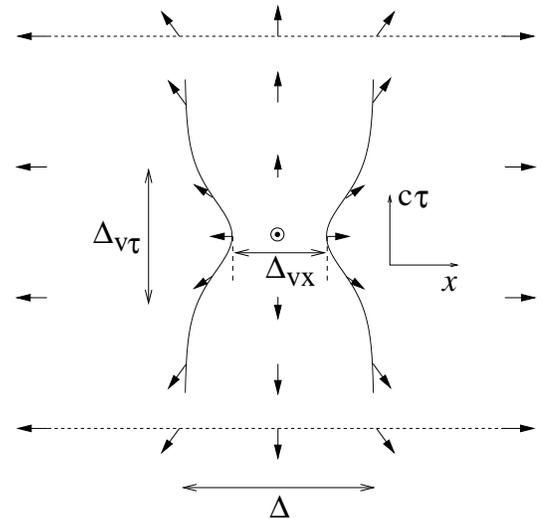,width=70mm,angle=-90.}}
\vspace{3mm}
  \caption{\label{fig:inst} The structure of instanton connecting two
  DW states with opposite chiralities.  Arrows denote projections of
  vector $\mvec{l}$ on the easy plane, and  on the thin solid line vector
  $\mvec{l}$ forms the angle of about $45^{\circ}$ with the easy axis.
 }
\end{figure}

{\em Chirality tunneling in absence of magnetic field.\/} Let us
consider first the case $\mvec{H}=0$. Tunneling
between the DW states with opposite chiralities can be studied using
the instanton formalism. Since the tunneling here occurs between two
{\em inhomogeneous\/} states, the corresponding instantons
are non-one-dimensional (space-time).
The structure of instanton
solution $\mvec{l}_{\text{inst}}(x,\tau)$ is shown in Fig.\
\ref{fig:inst}; it has the following asymptotic behavior:
\begin{eqnarray}
  \label{inst-prop}
&&  l_{X}\to\pm\sigma',\; x\to\pm\infty,\quad
  l_{Y}\to\mp\sigma,\;  x=0,\;\tau\to\pm\infty\nonumber\\
&&  l_{Z}\to p=\pm1,\; x\to 0,\;\tau\to 0\,,
\end{eqnarray}
note the appearance of another topological
charge $p=\pm1$.  Along any closed path in the Euclidean plane going
around (but far from) the instanton center  vector $\mvec{l}$ rotates
by the angle $2\pi \nu$ in the easy plane $XY$, where
$\nu=\sigma\sigma'=\pm1$. Thus, the instanton configuration has the
properties of an {\em out-of-plane magnetic vortex\/} (i.e., with
$l_{Z}\not=0$ in the center) and is characterized by two topological
charges \cite{affleck89rev}: vorticity $\nu$ and polarization $p$. The
instanton solution satisfies the equations
\begin{eqnarray}
  \label{inst-eq}
  &&\mvec{\nabla}^2\theta +\sin\theta\cos\theta
  [(\rho+\cos^2\varphi)/\Delta^2-(\mvec{\nabla}\varphi)^2] =0,\nonumber\\
  &&\mvec{\nabla}\cdot(\sin^2\theta\mvec{\nabla}\varphi)
  -(1/\Delta^2)\sin^2\theta \sin\varphi\cos\varphi=0,
\end{eqnarray}
where we have introduced angular variables as $l_{X}+il_{Y}=\sin\theta
e^{i\varphi}$, $l_{Z}=\cos\theta$, and
$\mvec{\nabla}=(\partial_{1},\partial_{2})$ denotes the Euclidean
gradient.

We are not able to find the exact solution of Eqs.\ (\ref{inst-eq}),
but the tunneling amplitude can be estimated from approximate
arguments.  One can readily observe that, in contrast to the
same problem for ferromagnet \cite{noteFM}, Eqs.\ (\ref{inst-eq}),
as well as their solutions, are {\em real\/.} Thus in absence
of magnetic field the real part of ${\cal A}_{E}$ is given by
$n_{\perp}SW/4$, and the imaginary part is completely
determined by the topological term $Q$ (note that $Q$ is a total
derivative and does not
contribute to the equations of motion).
Then the procedure of constructing the instanton solution can be viewed
as minimization of $W$; 
it should be remarked that the real part of the Euclidean
action for the instanton formally 
coincides with the energy of a vertical Bloch line in a 2d
DW, which is rather well studied \cite{BICG}.

Another observation is that in AF in absence of magnetic field the
translational DW motion in real space and its internal degree of
freedom (chirality) are completely uncoupled, in contrast to the
situation in ferromagnets \cite{BraunLoss96}; for $\mvec{H}\not=0$ this
coupling appears but becomes important only in strong field regime
\cite{ivkol97prb}.  Thus for weak fields one can calculate the
transition amplitude (\ref{split}) assuming that the DW coordinate in
real space is just fixed.

Further, for uniform boundary
conditions at infinity the quantity $Q$ determining
$\mbox{Im}\,{\cal A}_{E}$ 
can take only integer values, but in our case
$Q=-p\nu/2=\pm{1\over2}$ is half-integer, which is typical for
out-of-plane vortices (see, e.g., \cite{affleck89rev}). For the given
$\sigma'$ there are two instanton solutions with the same vorticity
$\nu$ and opposite polarizations $p$ which equally contribute to
$\mbox{Re} {\cal A}_{E}$ but have different signs of $\mbox{Im}\,{\cal
A}_{E}$. The tunnel splitting is given by
\begin{equation}
  \label{split}
  \Gamma=
C\hbar\omega_{l}\,(n_{\perp} \widetilde{W}S/4)^{1/2}
e^{-n_{\perp}\widetilde{W}S/4}\,|\cos\Phi|\,, 
\end{equation}
where
$\widetilde{W}=W[\mvec{l}_{\text{inst}}(x,\tau)]-W[\mvec{l}_{0}(x)]$ is
the difference of the real part of Euclidean action calculated on the
instanton solution and on a static DW without instanton,
$\omega_{l}=(c/\Delta)\sqrt{\rho}$ is the frequency of the
out-of-plane magnon localized at the DW playing the role of ``attempt
frequency,'' $C$ is a numerical constant, and $\Phi$ is the phase
determined by the $p$-dependent part of $\mbox{Im}\,{\cal A}_{E}$. In
the simplest model with $\mvec{H}=0$ considered so far $\Phi=\pi
n_{\perp} S$, and thus the tunneling amplitude vanishes when $S$ is
half-integer and $n_{\perp}$ is odd. In any mesoscopic sample with
weakly fluctuating cross-section the value of $n_{\perp}\gg1$ will be
randomly odd or even for different samples or different parts of the
same sample.  Thus the tunneling is forbidden for approximately one
half of domain walls in half-integer $S$ antiferromagnets. This is
somewhat similar to the case of half-integer $S$ nanoparticles with
non-compensated total spin $S_{\text{tot}}$ where the Kramers theorem
forbids tunneling for the particles with half-integer $S_{\text{tot}}$
\cite{Loss+92}.  Below we will see that in presence of external
magnetic field the situation is different: both integer and
half-integer $n_{\perp}$ contribute, but with different tunneling
rates.

One can observe that 
the problem  has three different length scales: the DW thickness
$\Delta$, the vortex core size in spatial direction
$\Delta_{vx}$, and the characteristic size of the core in the imaginary
time direction $\Delta_{v\tau}$.
For strong easy-plane type anisotropy, $\rho\gg 1$, the vortex core is
nearly  axially symmetric: up to distances 
$r\ll\Delta$ the anisotropy in the easy plane can be
neglected, and 
 the solution in the core 
reduces to the well-known case of
a usual vortex in an easy-plane magnet \cite{KosVorMan83}, with
$\theta=\theta_{0}(\xi)$, $\varphi=\nu\chi$, $\nu=\pm1$,
\begin{equation}
  \label{iso}
 d^{2}\theta_{0}/ d\xi^{2} +(1  -\nu^{2}/\xi^{2})
 \sin\theta_{0}\cos\theta_{0} =0\,,
\end{equation}
here $(r,\chi)$ are the polar coordinates in $(x_{1},x_{2})$ plane,
$r=(x_{1}^{2}+x_{2}^{2})^{1/2}$, $\chi=\arctan(x_{2}/x_{1})$, and
$\xi=r\sqrt{\rho}/\Delta$. Thus one has
$\Delta_{vx}=\Delta_{v\tau}=\Delta/\sqrt{\rho}\ll \Delta$, so the core
is isotropic and much smaller than the DW thickness.

In the opposite ``almost easy-axis'' case $\rho\ll 1$ the vortex 
core is strongly asymmetric: its spatial size $\Delta_{vx}$ 
coincides with the DW
thickness $\Delta$, but the imaginary time size
$\Delta_{v\tau}=\Delta/\sqrt{\rho}$ is much larger \cite{comment}.

On the other hand, for $r\gg \Delta_{vx}, \Delta_{v\tau}$, i.e., far
outside the core, one can put $\theta\simeq \pi/2$, which reduces the
system (\ref{inst-eq}) to the 2d elliptic sine-Gordon equation,
\begin{equation}
  \label{far}
 \mvec{\nabla}^{2}\varphi=(1/2\Delta^{2})\sin2\varphi\,.
\end{equation}

In the large rhombicity limit $\rho\gg1$ within a wide range of $r$
(for $\Delta/\sqrt{\rho}\ll r\ll \Delta$) the solutions (\ref{iso})
and (\ref{far}) can be regarded as coinciding, and the integrand in
$\widetilde{W}$ is proportional to $1/r^{2}$.  Then, one may divide
the integration domain into two parts: $r<R$ and $r>R$, where $R$ is
an arbitrary point in between $\Delta/\sqrt{\rho}$ and $\Delta$. For
$r<R$ the solution (\ref{iso}) may be used, yielding
$\widetilde{W}_{r<R}=2\pi\ln(\zeta R\sqrt{\rho}/\Delta)$ with the
numerical factor $\zeta$ being known from
Ref. \onlinecite{KosVorMan83}, $\zeta\simeq4.2$.  For $r>R$, one can
use a trial function approximately satisfying (\ref{far}), e.g.,
\begin{eqnarray}
  \label{trial}
&&\tan \varphi=\tanh[x_{2}/(\Delta_{vx}^{-1}\Delta_{v\tau}\Delta)] 
/\sinh(x_{1}/\Delta)\,,\nonumber\\
&&\cos\theta=[\cosh(x_{1}/\Delta_{vx})\cosh(x_{2}/\Delta_{v\tau})]^{-1}\,,
\end{eqnarray}
and evaluate the integral in $\widetilde{W}$ numerically, which for
$\rho\gg1$ gives $\widetilde{W}_{r>R}=2\pi\ln(\zeta'\Delta/R)$ with
$\zeta'\simeq0.525$. Summing up the two contributions, we obtain
\begin{equation}
  \label{W-largerho}
  \widetilde{W}\simeq 2\pi\ln(2.2 \sqrt{\rho}),\quad \rho\gg1\,.
\end{equation}

In the weak rhombicity limit $\rho\ll 1$ the trial function
(\ref{trial}) can be used for the entire $(x_{1},x_{2})$ plane, yielding
the result
\begin{equation}
  \label{W-lowrho}
\widetilde{W}\simeq 8\rho^{1/2},\quad  \rho\ll1\,,
\end{equation}
which coincides with one obtained by us earlier in the effective
Lagrangian approach\cite{ivkol96jetp}. The result
(\ref{W-lowrho}) breaks down only for extremely small
rhombicities $\rho\ll 4/(n_{\perp}S)^{2}$, when the system is very
close to the easy-axis regime and its low-energy spectrum coincides
with that of the free rotator, which yields the tunnel splitting
$\Gamma\sim (\hbar c/2\Delta n_{\perp}S)$ \cite{ivkol96jetp}.

Comparing the results for tunneling in AF domain wall
with those for a
ferromagnetic DW \cite{BraunLoss96}, one can see that (i) for a
ferromagnetic DW the function $\widetilde{W}$ determining the
tunneling exponent contains the additional large factor $\Delta/a$;
(ii) for ferromagnet $\widetilde{W}\propto\sqrt{\rho}$ at large $\rho$
while for AF the growth of $\widetilde{W}(\rho)$ is much slower.

{\em Effect of magnetic field.\/} Consider now the behavior of the
imaginary part of the Euclidean action when a weak external magnetic
field $\mvec{H}$ is applied to the system (we neglect here the
anisotropy renormalization $\widetilde{w}_{a}$ because its effect is 
rather trivial). 
One can see that the mixed
product in $Q_{H}'$ significantly differs from zero  only in the vortex core,
and thus 
in case of large rhombicity $\rho\gg1$
the isotropic vortex solution (\ref{iso}) may be used to
estimate it. After integration we obtain
\begin{equation}
  \label{addQ}
 Q'_{H}\approx p\lambda (H_{X}/H_{c}) \,,
\end{equation}
where $H_{c}=(4Z_{c}S^{2}JK_{1})^{1/2}/\gamma$ denotes the magnitude of field for which the
field-induced anisotropy becomes equal to the easy-plane one, 
$H_{X}$ is the field
component along the easy axis,
$\lambda=\int_{0}^{\infty}d\xi\{{1\over2}\sin2\theta_{0}
+\xi(d\theta_{0}/d\xi)\}$ is a numerical constant, $\lambda\approx
3.83$.  This  results in the following expression for the phase factor
$\Phi$ in (\ref{split}):
\begin{equation} 
\label{resH} 
\Phi\mapsto\Phi_{H}=\pi n_{\perp}S [1+(\lambda H_{X}/2H_{c}] \,,
\end{equation}
thus the tunneling amplitude oscillates
as a function of the field with the orientation-dependent period 
\begin{equation} 
\label{period} 
\delta H=(2H_{c}/ n_{\perp} S \lambda)\,.
\end{equation}
This period can be rather small: assuming $S={5\over2}$ and a typical
$H_{c}\sim 100$~kOe, one gets $\delta H\sim 2$~kOe$\div20$~Oe for
$n_{\perp}=10\div 10^{3}$.  A similar oscillating behavior was
predicted earlier for tunneling in small ferromagnetic
\cite{BogachekKrive92} and antiferromagnetic
\cite{GolyshevPopkov95,ChioleroLoss97} particles, with the difference
that in the AF case instead of the field $H_{c}\propto \sqrt{JK_{1}}$
in (\ref{period}) a much stronger exchange field $H_{e}\propto J$
would be present.  For half-integer $S$
presence of the magnetic field lifts the degeneracy of two
odd-$n_{\perp}$ DW states
with opposite chirality, allowing tunneling between them.  Note that
the ``correction'' in $\Phi$ is  strong even for weak $H$ since
it contains a large factor $n_{\perp}$.

Another consequence of the above result is that {\em for half-integer
$S$ there are two different values of the tunnel splitting for even
and odd $n_{\perp}$,\/} which means that for any experimental setup
with weakly fluctuating $n_{\perp}$ there should be {\em two\/}
different resonance peaks at the corresponding frequencies; this
effect was overlooked in previous studies.  It is worthwhile to remark
that the same effect should be also present in half-integer $S$ AF
nanoparticles with noncompensated spins considered first by Loss {\em
et al.}\cite{Loss+92} and also studied later in Refs.\
\onlinecite{Chud+95}, provided that there is some weak interaction
(e.g., magnetic field or the Dzyaloshinskii-Moriya interaction
\cite{iknato97}) contributing to the phase factor $\Phi$ and shifting
it from a multiple of ${\pi\over2}$.

We would like to finish with a word of caution: in presence of field
the problem of chirality tunneling is actually more complicated then
one can guess from the simple arguments presented above.  The point is
that the field contribution $Q_{H}'$, unlike $Q$, is {\em not\/} a
total derivative and thus yields an {\em imaginary} perturbation to
the equations of motion, causing nontrivial changes in the instanton
structure and in the spectrum of eigenmodes in presence of instanton
which eventually contribute to the phase factor $\Phi$
\cite{ChioleroLoss97}.  However, using a perturbation theory in
$\mvec{H}$, one can show that corrections from the change of instanton
structure contribute to $\widetilde{W}$ as $(H/H_{c})^{2}$ and to
$\Phi$ as $n_{\perp} (H/H_{c})^{3}$, so that they can be neglected for
weak fields, and the contribution to $\Phi$ from fluctuations does not
contain $n_{\perp}$. Thus we think that the formula (\ref{period})
remains correct, together with our main conclusions on the oscillating
field dependence of $\Gamma$ and on presence of two different
tunneling rates.

{\em Acknowledgements.\/} This work was supported in part by the
Ukrainian Ministry of Science (grant 2.4/27) and was finished during
the stay of B.I. in Hannover Institute for Theoretical Physics,
supported by Deutsche Forschungsgemeinschaft. A.K. was supported by
the German Ministry for Research and Technology (BMBF) under the
contract 03MI4HAN8.

\vspace{3mm}

\begin{references}

\bibitem{QTM} See for a review {\em Quantum Tunneling of
Magnetization,\/} ed. by L. Gunther and B. Barbara, NATO ASI Ser. E,
vol. 301 (Kluwer, Dordrecht, 1995).

\bibitem{Gatteschi+94} D. Gatteschi, A. Caneschi, L. Pardi, and
R. Sessoli, Science {\bf 265}, 1054 (1994).

\bibitem{ivkol94+GalkinaIvanov95} B. A. Ivanov and A. K. Kolezhuk.
JETP Lett. {\bf 60}, 805 (1994);
E. G. Galkina and B. A. Ivanov, JETP Lett. {\bf 61}, 511 (1995).

\bibitem{BraunLoss96} H.-B. Braun and D. Loss, Phys. Rev. B {\bf 53},
3237 (1996); Int. J. Mod. Phys. B {\bf 10}, 219 (1996).

\bibitem{BarbaraChud+KriveZasl90} B. Barbara and E. M. Chudnovsky,
Phys. Lett. {\bf A145}, 205 (1990); I. Krive and O. B. Zaslavskii,
J. Phys.: Condens. Matter {\bf 2}, 9457 (1990).

\bibitem{fradkin}  E. Fradkin, {\em Field theories of condensed
matter systems}, in {\em Frontiers in Physics} vol. 82, Addison-Wesley
(1991). 

\bibitem{Berry84}  M. V. Berry, Proc. Roy. Soc. London A {\bf
232}, 45 (1984).

\bibitem{affleck89rev} I. Affleck, J. Phys. Condens. Matter {\bf
1}, 3047 (1989).

\bibitem{noteFM} In ferromagnet $\cos\theta$ and $\varphi$ are
conjugate variables, thus the equations of motion contain first-order
time derivatives, and $\theta,\varphi$ in the instanton solution
become complex, with real and imaginary parts being of the same order
of magnitude.


\bibitem{BICG}
 V. G. Bar'yakhtar, M. V. Chetkin, B. A. Ivanov, and
S. N. Gadetskii, {\em Dynamics of topological magnetic solitons.
Experiment and Theory.\/} -- Springer Tracts in Modern Physics,
vol.~129, Springer-Verlag, 1994.

\bibitem{ivkol97prb} B.A. Ivanov and A.K. Kolezhuk, Phys. Rev. B {\bf
56}, 8886 (1997).

\bibitem{Loss+92} D. Loss, D. P. DiVincenzo, and
G. Grinstein, Phys. Rev. Lett. {\bf 69}, 3232 (1992);  J. von
Delft and C. L. Henley, Phys. Rev. Lett. {\bf 69}, 3236 (1992).


\bibitem{KosVorMan83} A. M. Kosevich, V. P. Voronov, and I. V. Manzhos,
  Sov. Phys. JETP {\bf 57}, 86 (1983). 


\bibitem{comment} It is most convenient to see that with a different
choice of angular variables, $l_{X}=\cos\Theta$,
$l_{Y}+il_{Z}=\sin\Theta\cos\Phi$, then the characteristic scale of
$\Phi$ variation $\Delta/\sqrt{\rho}$ is much larger than the scale of
$\Theta$ variation which is just $\Delta$.


\bibitem{ivkol96jetp} B. A. Ivanov and A. K. Kolezhuk,
Phys. Rev. Lett. {\bf 74}, 1859 (1995); JETP {\bf
83}, 1202 (1996).

\bibitem{BogachekKrive92} E. N. Bogachek and I. V. Krive, Phys. Rev. B
{\bf 46}, 14559 (1992).

\bibitem{GolyshevPopkov95} V. Yu. Golyshev and A. F. Popkov,
Europhys. Lett. {\bf 29}, 327 (1995).

\bibitem{ChioleroLoss97} A. Chiolero and D. Loss,
Phys. Rev. Lett. {\bf 80}, 169 (1998).

\bibitem{Chud+95} E. M. Chudnovsky, J. Magn. Magn. Mater. {\bf
140-144}, 1821 (1995); J. M. Duan and A. Garg, J. Phys.:
Condens. Matter {\bf 7}, 2171 (1995);  A. Chiolero and D. Loss, PRB
{\bf 56}, 738 (1997).

\bibitem{iknato97}  B. A. Ivanov and A. K. Kolezhuk, e-print cond-mat/9706292.


\end{references}
\end{document}